\documentclass[aps,pra,reprint, amsmath, amssymb,superscriptaddress,nofootinbib]{revtex4-1}

\usepackage{bbold}
\usepackage{bm}
\usepackage[retainorgcmds]{IEEEtrantools}
\usepackage{graphicx}
\usepackage{mathrsfs}
\usepackage{amsmath}
\usepackage{amssymb}
\usepackage{color,xcolor}
\usepackage{dsfont,amsfonts}
\usepackage{times,txfonts}
\usepackage{nicefrac}
\usepackage[colorlinks=true,linkcolor=blue,urlcolor=blue,citecolor=blue,pdfusetitle]{hyperref}
\usepackage{soul}
\usepackage{lipsum}

\newcommand{\proj}[1]{|#1\rangle\langle #1|}
\newcommand{\ket}[1]{|#1\rangle}
\newcommand{\bra}[1]{\langle#1|}

\newcommand{\tr}{\text{Tr}}

\begin{document}

\title{Joint fluctuation theorems for sequential heat exchange} 
\date{\today}
\author{Jader P. Santos}
\affiliation{Instituto de F\'isica da Universidade de S\~ao Paulo,  05314-970 S\~ao Paulo, Brazil.}
\author{Andr\'e M. Timpanaro}
\affiliation{Universidade Federal do ABC,  09210-580 Santo Andr\'e, Brazil}
\author{Gabriel T. Landi}
\email{gtlandi@if.usp.br}
\affiliation{Instituto de F\'isica da Universidade de S\~ao Paulo,  05314-970 S\~ao Paulo, Brazil.}

\begin{abstract}
We study the statistics of heat exchange of a quantum system that collides sequentially with an arbitrary number of ancillas.
This can describe, for instance, an accelerated particle going through a bubble chamber. 
Unlike other approaches in the literature, our focus is on the \emph{joint} probability distribution that heat $Q_1$ is exchanged with ancilla 1, heat $Q_2$ is exchanged with ancilla 2, and so on. 
This  allows one to address questions concerning the correlations between  the collisional events. 
The joint distribution is found to satisfy a Fluctuation theorem of the Jarzynski-W\'ojcik type. 
Rather surprisingly, this fluctuation theorem links the statistics of multiple collisions  with that of independent single collisions, even though the heat exchanges are statistically correlated.


\end{abstract}
\maketitle{}

\section{Introduction}

Fluctuations of thermodynamic quantities, which are usually negligible in macroscopic systems, are known to play a dominant role in the micro- and mesoscopic domain. 
These fluctuations are embodied in the so-called  fluctuation theorems (FT)~\cite{Evans1993,Gallavotti1995b,Jarzynski1997,Crooks1999}, a collection of predictions for systems evolving under nonequilibrium conditions valid beyond linear response.
They can be summarized as~\cite{Campisi2011,Esposito2009}
\begin{align} \label{eq:FT}
\frac{P(+\Sigma)}{\tilde{P}(-\Sigma)} = e^\Sigma,
\end{align}
where $P(\Sigma)$ denotes the probability that an amount of entropy $\Sigma$ is produced in a certain process and $\tilde{P}(\Sigma)$ denotes the corresponding probability for the time-reversed process.

Of the many scenarios which present FTs, one which is particularly interesting is that of heat exchange between a system $S$, prepared in equilibrium with a temperature $T_s$, and an environment $E$, prepared in a different temperature $T_e$. 
In this case, as first shown by Jarzynski and W\'ojcik in Ref.~\cite{Jarzynski2004}, the distribution $P(Q)$, of the heat exchanged between them, satisfies 
\begin{align} \label{eq:XFT}
\frac{P(+Q)}{\tilde{P}(-Q)} = e^{\Delta\beta   Q},
\end{align}
where $\Delta\beta = \beta_e-\beta_s$ (with $\beta = 1/T$ and $k_B = 1$). 
Quite surprising, in this case it turns out that $\tilde{P}(Q) = P(Q)$, meaning the statistics of the forward and backward processes are the same. 
Eq.~(\ref{eq:XFT}) was subsequently generalized  to allow for the exchange of both energy and particles between several interacting systems initially at different temperatures and chemical potentials \cite{Saito2008, Andrieux2006a,Esposito2009}.


Here we consider a generalization of this scenario, where the system interacts sequentially with multiple parts of the environment, exchanging a small amount of heat with each part. 
One can imagine, for instance, an accelerated particle crossing a bubble chamber. 
In this case,  the system will leave a trail on $E$, represented by the heat exchanged in each point. 
In the microscopic domain this process will be stochastic, with a random amount of heat exchanged in each interaction. 

The key idea that we will explore in this paper is to look at the \emph{joint probability distribution} for the heat exchanged with each part, $P(Q_1, Q_2, Q_3, \ldots)$. 
This allows us to understand the correlations between the different heat exchanges.
For instance, from a stochastic perspective a large  exchange in the first collision increases the probability that the second collision exchanges less. 
This feature is fully captured by the joint distribution. 

To formalize this idea, we split the environment into a set of ancillas $A_i$, with which the system interacts sequentially,  producing a collisional model~\cite{Rodrigues2019,Strasberga, DeChiara2018, Scully2003}. 
The process is schematically illustrated in FIG.~\ref{fig:forward} and the formal framework is developed in Sec.~\ref{sec:framework}. 
In Sec.~\ref{sec:FT} we then show that $P(Q_1, Q_2, Q_3, \ldots)$ satisfies a fluctuation theorem that generalizes~(\ref{eq:XFT}).
Moreover, we show how this fluctuation theorem relates the joint distribution to the statistics of a single collision, even though the events are statistically correlated. 




\section{\label{sec:framework}Formal framework}

\begin{figure}
\centering
\includegraphics[width=0.4\textwidth]{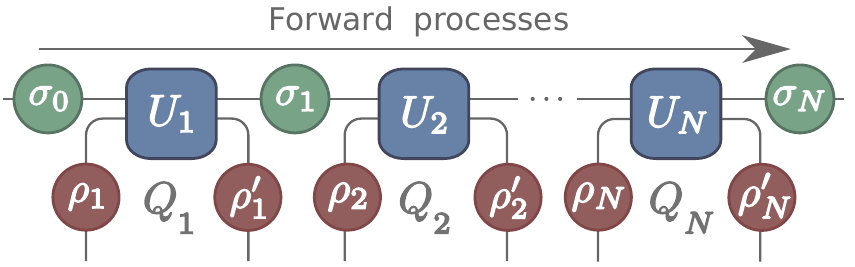}
\caption{\label{fig:forward} Schematic representation of a system $S$ interacting sequentially with a series of ancillas.
The system starts in the state $\sigma_0$ and the ancillas in an initial states $\rho_i$, which are assumed to be thermal but at possibly different temperatures.
Each $SA_i$ interaction is also governed by a possibly different unitary $U_i$. 
}
\end{figure}

We consider a quantum system $S$, with Hamiltonian $H^s$, prepared in a thermal state 
$\sigma_0=e^{-\beta_s H^s}/Z_s$, with temperature $T_s$.
The system is  put to interact sequentially  with a series of $N$ ancillas $A_i$, as depicted in  FIG.~\ref{fig:forward}. 
The ancillas are not necessarily identical. 
Each has Hamiltonian $H^i$ and is prepared in a  thermal state $\rho_i=e^{-\beta_i H^i}/Z_i$,  with possibly different temperatures $T_i$. 
Each collision is described by a unitary operator $U_i$ acting only on $SA_i$, which may also differ from one interaction to another. 

In order to comply with the scenario of Ref.~\cite{Jarzynski2004}, we assume that the $U_i$ satisfy the strong energy-preservation condition
\begin{align}\label{energy_preservation}
[U_i,H^s+H^i]=0.  
\end{align}
Or, what is equivalent, that each collision is a  thermal operation~\cite{Brandao2013,Brandao2015}.
This implies that all  energy that leaves $S$ enters $A_i$, so nothing is stuck in the interaction. 
As a consequence, there is no work involved and all the change in energy of the system can be unambiguously identified as heat flowing to the ancillas~\cite{DeChiara2018}.

We label the eigenvalues and eigenvectors of the system as 
$H^s\ket{\alpha}=E_{\alpha}^s\ket{\alpha}$. 
For concreteness, we assume these levels are non-degenerate. 
Time is labeled discretely by $i = 1,2,3,\ldots$, representing which collisions already took place. 
For instance, the initial state is  decomposed as $\sigma_0 = \sum_{\alpha_0} p_0(\alpha_0) |\alpha_0 \rangle \langle \alpha_0|$, with $p_0(\alpha_0) = e^{- \beta_s E_{\alpha_0}^s}/Z_s$ and we use $\alpha_0$ to emphasize that this is before the first collision. 
Similarly, the eigenvalues and eigenvectors of the ancillas are labeled as $H_i |n_i \rangle = E_{n_i}^e |n_i\rangle$. 
The initial state of each $A_i$ is thus decomposed as  $\rho_i=\sum_{n_i}q_i(n_i)\proj{n_i}$ where $q_i(n_i)=e^{-\beta_i E_{n_i}^e}/Z_i$.

The dynamics depicted in FIG.~\ref{fig:forward} generates a stroboscopic map for the system. 
The joint state of $SA_i$ after the interaction is given by 
\begin{equation}\label{varrho}
    \varrho_i = U_i \big(\sigma_{i-1}\otimes\rho_i\big)U^\dagger_i. 
\end{equation}
Taking the partial trace over $A_i$ then leads to the updated state $\sigma_i$. 
Conversely, tracing over the system leads to the reduced state $\rho_i'$ of the ancilla after the interaction (FIG.~\ref{fig:forward}).

The fact that the unitary is energy preserving [Eq.~(\ref{energy_preservation})], together with the assumption that the energy levels are non-degenerate, mean that it is possible to construct  quantum trajectories for the system in two equivalent ways. 
The first is to assume a two-point measurement scheme in $S$ at each step~\cite{Campisi2010a,Utsumi2010}. 
Eq.~(\ref{energy_preservation}) implies that the system will remain  diagonal in the energy basis, so that measurements in this basis are non-invasive (that is, have no additional entropy production associated to it). 
Measuring $S$ in the energy basis after each collision then leads to the trajectory 
\begin{equation} \label{eq:traje_s}
\gamma_s=\{\alpha_0,\alpha_1,\dots,\alpha_N\}.
\end{equation}
The heat associated with each collision is then readily defined as 
\begin{equation}\label{heat_def}
    Q_i[\gamma_s] = -E_{\alpha_i}^s + E_{\alpha_{i-1}}^s, 
\end{equation}
which we label to be positive when energy leaves the system. 

Alternatively, one can construct a quantum trajectory by measuring the ancillas, before and after each collision, plus a single measurement of the system before the process starts.
That is, one can  consider instead a quantum trajectory of the form 
\begin{equation}\label{eq:traj_e}
\gamma_e=\{\alpha_0,n_1,n_1',n_2,n_2',\dots,n_N,n_N'\}.
\end{equation}
This, in a sense, is much more natural since the ancillas are only used once and thus may be experimentally more easily accessible. 
And as far as heat exchange is concerned, this turns out to be  equivalent to the trajectory~(\ref{eq:traje_s}). 
The reason is that
Eq.~(\ref{energy_preservation}) implies the restriction 
\begin{equation}\label{energy_preservation_delta}
    \langle \alpha_i n_i' | U_i | \alpha_{i-1} n_i\rangle \propto \delta\big(E_{\alpha_i}^s + E_{n_i'}^e = E_{\alpha_{i-1}}^s + E_{n_i}^e\big),
\end{equation}
where $\delta(a=b)$ is the Kronecker delta. 
In addition, since the energy values are taken to be non-degenerate, energies uniquely label states. 
Thus, for instance, if we know $\alpha_0, n_1, n_1'$ we can  uniquely determine $\alpha_1$, and so on. 
The converse, however, is not true: from $\alpha_0$ and $\alpha_1$ we cannot specify $n_1$ and $n_1'$ (which is somewhat evident given that the number of points in Eq.~(\ref{eq:traje_s}) is smaller than that in Eq.~(\ref{eq:traj_e})). 
This, however, is not a problem if one is interested only in the heat exchanged, which can also be defined from the trajectory~(\ref{eq:traj_e}) as
\begin{equation}\label{heat_def_env}
    Q_i[\gamma_e] = E_{n_i'}^e - E_{n_i}^e.
\end{equation}
Due to Eq.~(\ref{energy_preservation_delta}) this must coincide with Eq.~(\ref{heat_def}); i.e., $Q_i[\gamma_e] \equiv Q_i[\gamma_s]$.

The assumption in Eq.~(\ref{energy_preservation}) may at first seem somewhat artificial. 
But this is not the case. 
This assumption is a way to bypass the idea of weak coupling, which is one of the conditions used in~\cite{Jarzynski2004}. 
It can be viewed as a kind of ``weak coupling \emph{a priori}''; that is, instead of using weak coupling as an approximation, we impose it from the start as an assumption. 
Moreover, the interesting thing about the present analysis is that it establishes under which conditions Eqs.~(\ref{eq:traje_s}) and~(\ref{eq:traj_e}) are equivalent. Naively one would expect that this is often the case. But, as the above arguments show, several assumptions are necessary for this to be the case. 
This reflects some of the challenges that appear in describing thermodynamics in the quantum regime.

\subsection{Path probabilities from measurements in $S$}

Thermal operations imply that the probability that, after the $i$-th collision, the system is in a given eigenstate $|\alpha_i\rangle$ depends only on the probabilities in the previous time. 
That is, the dynamics of populations and coherences completely decouple~\cite{Santos2017a}. 
Indeed, Eq.~(\ref{varrho}) together with Eq.~(\ref{energy_preservation}) imply that 
\begin{align} \label{eq:p_i}
p_i(\alpha_i) &= \bra{\alpha_i}\sigma_i\ket{\alpha_i} =\sum_{\alpha_{i-1}} M_i(\alpha_i|\alpha_{i-1})p_{i-1}(\alpha_{i-1}),
\end{align}
where 
\begin{align} \label{eq:propagator}
M_i(\alpha_i|\alpha_{i-1}) = \sum_{n_i,n_i'} 
|\bra{\alpha_i,n_i'}U_i\ket{\alpha_{i-1},n_i}|^2 q_i(n_i).
\end{align}
The populations therefore evolve as a classical Markov chain, with $M_i(\alpha_i|\alpha_{i-1})$ representing the  transition probability of going from $\alpha_{i-1}$ to $\alpha_i$. 
Moreover, Eq.~(\ref{energy_preservation_delta}) together with the fact that the ancillas are initially thermal, imply that $M_i(\alpha_i|\alpha_{i-1})$ satisfies detailed balance 
\begin{equation} \label{eq:DBR}
M_i(\alpha_i|\alpha_{i-1}) 
e^{-\beta_i E^s_{\alpha_{i-1}^{}}}
= 
M_i(\alpha_{i-1}|\alpha_i)
e^{-\beta_i E^s_{\alpha_{i}^{}}},
\end{equation}
where, notice, what appears here is the temperature $\beta_i$ of ancilla $A_i$.

The path probability associated with $\gamma_s$ in Eq.~(\ref{eq:traje_s}) will then be 
\begin{equation} \label{eq:Ps}
\mathcal{P}[\gamma_s] = 
M_N(\alpha_N|\alpha_{N-1})\dots 
M_2(\alpha_2|\alpha_{1})M_1(\alpha_1|\alpha_{0})p_0(\alpha_0),
\end{equation}
which is nothing but the joint distribution of a Markov chain. 
We call attention to the clear causal structure of this expression: marginalizing over future events has no influence on past ones. For instance, summing over $\alpha_N$ leads to a distribution of the exact same form. 
Conversely, marginalizing over past variables completely changes the distribution.

The joint distribution of heat can then be constructed from Eq.~(\ref{eq:Ps}) in the usual way:
\begin{equation}\label{PQ}
    P(Q_1,\ldots,Q_N) = \sum\limits_{\gamma_s} \mathcal{P}[\gamma_s] \bigg(\prod\limits_{i=1}^N \delta\big(Q_i - Q_i[\gamma_s]\big)\bigg). 
\end{equation}
This is the basic object that we will explore in this paper. 

\begin{widetext}
\subsection{Path probabilities from measurements in the $A_i$}

Alternatively, we also wish to show how Eq.~(\ref{PQ}) can be constructed from the trajectory $\gamma_e$ in Eq.~(\ref{eq:traj_e}). 
The easiest way to accomplish this is to first consider the augmented trajectory
\begin{equation}\label{gamma_se}
\gamma_{se} = \{\alpha_0, n_1,n_1',\alpha_1,n_2,n_2',\alpha_2,\ldots,n_N,n_N',\alpha_N\}
\end{equation}
Introducing the transition probabilities $R_i(\alpha_i, n_i'| \alpha_{i-1},n_i) = |\bra{\alpha_i,n_i'}U_i\ket{\alpha_{i-1},n_i}|^2$, the path distribution associated with the augmented trajectory $\gamma_{se}$ will be 
\[
\mathcal{P}[\gamma_{se}] =  R_N(\alpha_N, n_N'|\alpha_{N-1},n_N)\ldots R_1(\alpha_1,n_1'|\alpha_0,n_1) q_N(n_N)\ldots q_1(n_1) p_0(\alpha_0).
\]
As a sanity check, if we marginalize this over $n_i$ and $n_i'$ we find 
\begin{IEEEeqnarray*}{rCl}
\mathcal{P}[\gamma_s] &=& \sum\limits_{\substack{n_1,\ldots,n_N \\ n_1',\ldots, n_N'}} R_N(\alpha_N, n_N'|\alpha_{N-1},n_N)\ldots R_1(\alpha_1,n_1'|\alpha_0,n_1) q_N(n_N)\ldots q_1(n_1) p_0(\alpha_0)
\\[0.2cm]
&=&M_N(\alpha_N|\alpha_{N-1})\dots 
M_2(\alpha_2|\alpha_{1})M_1(\alpha_1|\alpha_{0})p_0(\alpha_0),
\end{IEEEeqnarray*}
where we used Eq.~(\ref{eq:propagator}). This is therefore precisely  $\mathcal{P}[\gamma_s]$ in Eq.~(\ref{eq:Ps}), as expected. 

Instead, from $\mathcal{P}[\gamma_{se}]$ one can now obtain  $\mathcal{P}[\gamma_e]$  by marginalizing over $\alpha_1,\ldots, \alpha_N$; viz.,
\begin{equation}
\mathcal{P}[\gamma_{e}] =  \sum\limits_{\alpha_1,\ldots,\alpha_N} R_N(\alpha_N, n_N'|\alpha_{N-1},n_N)\ldots R_1(\alpha_1,n_1'|\alpha_0,n_1) q_N(n_N)\ldots q_1(n_1) p_0(\alpha_0).
\end{equation}
\end{widetext}
The above analysis puts in evidence the Hidden Markov nature of the dynamics in FIG.~\ref{fig:forward}. 
When measurements are done in the ancilla, the system plays the role of the hidden layer, which is not directly accessible. Instead, predictions about the system must be made from the visible layer (i.e., the ancillas). 

This Hidden Markov nature manifests itself on the fact that even though the system obeys a Markov chain [Eq.~(\ref{eq:Ps})], the same is \emph{not} true for the ancillas. 
In symbols, this is manifested by the fact that $n_i'$ depends not only on $n_i$ and $n_{i-1}'$, but on the entire past history $(n_1,n_1',\ldots,n_{i-1},n_{i-1}',n_i)$. 
This is intuitive in a certain sense: the amount of heat exchanged at the $i$-th collision will depend on the heat exchanged in all past events.

With $\mathcal{P}[\gamma_e]$, the distribution of heat, Eq.~(\ref{PQ}) can be equivalently defined using Eq.~(\ref{heat_def_env}). One then finds  
\begin{equation}\label{PQ_env}
    P(Q_1,\ldots,Q_N) = \sum\limits_{\gamma_e} \mathcal{P}[\gamma_e] \bigg(\prod\limits_{i=1}^N \delta\big(Q_i - Q_i[\gamma_e]\big)\bigg). 
\end{equation}
The reason why this is equivalent to Eq.~(\ref{PQ})  becomes clear from the way we derived $\mathcal{P}[\gamma_e]$ above: we can expand the summation to $\gamma_{se}$ and then use the fact that $Q_i[\gamma_s] = Q_i[\gamma_e]$.

\subsection{\label{sec:back}Backward process}

\begin{figure}
\centering
\includegraphics[width=0.4\textwidth]{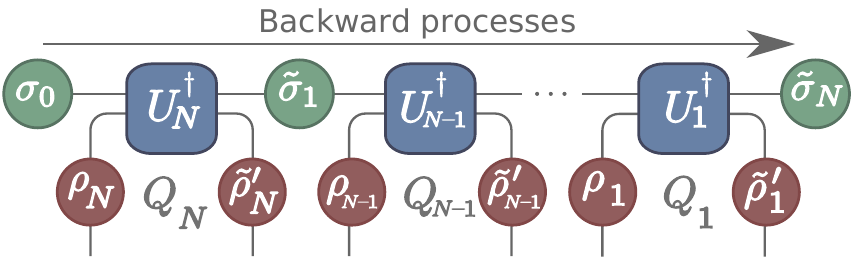}
\caption{\label{fig:backward} Schematic representation of the  backward process.}
\end{figure}

To construct the fluctuation theorem, we must now establish the backward process. 
As shown in~\cite{Manzano2018}, however,  there is an arbitrariness in the choice of the initial state of the backward process; different choices lead to different definitions of the entropy production. 
Here we are interested specifically in heat and the generalization of the Jarzynski-W\'ojcik fluctuation theorem~\cite{Jarzynski2004}. 
Hence, we assume that in the backward process both system and ancillas are fully reset back to their thermal states. 
As usual, the time-reversed interaction between $SA_i$ now takes place by means of the unitary $U_i^\dagger$. 
However,  the order of the interactions must now be flipped around, as shown in FIG.~\ref{fig:backward}.




In the backward process, the system will therefore evolve according to
\[
\tilde{p}_i(\alpha_{N-i}) = \sum_{\alpha_{N-i+1}} M_{N-i+1}(\alpha_{N-i}|\alpha_{N-i+1})\tilde{p}_{i-1}(\alpha_{N-i+1}),
\]
where we index the states as $\alpha_{N-i}$ instead of $\alpha_i$ just so that the trajectory $\gamma_s$ can remain the same as in the forward process. 
The path probability $\tilde{\mathcal{P}}[\gamma_s]$ associated to this process will then be 
\begin{equation} \label{eq:tildePs}
\tilde{\mathcal{P}}[\gamma_{s}] = 
M_1(\alpha_0|\alpha_1)\dots 
M_N(\alpha_{N-1}|\alpha_N)p_0(\alpha_N),
\end{equation}
which is similar to that used in the original Crooks fluctuation theorem~\cite{Crooks1998}. 
The corresponding heat distribution is
\begin{equation}\label{PQ_back}
    \tilde{P}(Q_N,\ldots,Q_1) = \sum\limits_{\gamma_s} \tilde{\mathcal{P}}[\gamma_s] \prod\limits_{i=1}^N \delta\big(Q_i + Q_i[\gamma_s]\big),
\end{equation}
where $Q_i$ continues to be the heat exchanged with $A_i$ (which is now different from the heat exchanged at collision $i$).

\section{\label{sec:FT}Joint fluctuation theorem for heat exchange}

We are now ready to construct the fluctuation theorem. 
The detailed balance condition~(\ref{eq:DBR}) immediately implies that Eqs.~(\ref{PQ}) and (\ref{PQ_back}) will be related by 
\begin{equation}\label{FT}
    \frac{P(Q_1,\ldots,Q_N)}{\tilde{P}(-Q_N, \ldots, -Q_1)} = e^{\sum_{i=1}^N (\beta_i - \beta_s) Q_i}. 
\end{equation}
This is a theorem for the joint distribution of the heat exchanged between multiple ancillas. It thus represents a generalization of Ref.~\cite{Jarzynski2004} to the case where the system interacts sequentially with multiple reservoirs. 
This result has several features which are noteworthy. 
First, note that the temperature $\beta_i$ of the ancillas are not necessarily the same. 
Second, note how after the first collision the state of the system is no longer thermal. But still, this does not affect the fluctuation theorem. All that matters is that before the first collision the system is in equilibrium. 

\subsection{Causal order and relation to single collisions}

The causal order of the process plays a crucial role here. 
Marginalizing over future events has no effect on the fluctuation theorem. 
That is, from~(\ref{FT}) one could very well construct a similar relation for $P(Q_1,\ldots, Q_{N-1})$, by simply summing over $Q_N$. 
This is not possible, however, for marginalization over past events. 
That is, $P(Q_2,\ldots, Q_N)$, for instance, does \emph{not} satisfy a fluctuation theorem. 

\begin{figure}
\centering
\includegraphics[width=0.4\textwidth]{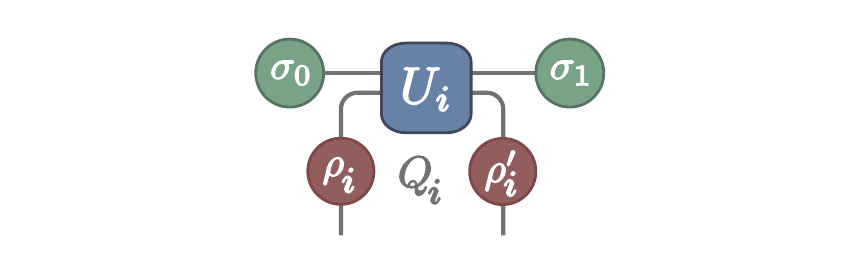}
\caption{\label{fig:single} Schematic representation of a single collision event.}
\end{figure}

The right-hand side of Eq.~(\ref{FT}) is very similar to what appears in the original FT~(\ref{eq:XFT}). 
We can make this more rigorous as follows. 
Let us consider a \emph{different} process, consisting of a single collision between the system thermalized in $\beta_s$ and an ancilla thermalized in $\beta_i$ (FIG.~\ref{fig:single}). 
The associated heat distribution $P_\text{sc}(Q_i)$ will then satisfy Eq.~(\ref{eq:XFT}); viz., 
\begin{equation}
    \frac{P_\text{sc}(Q_i)}{P_\text{sc}(-Q_i)} = e^{(\beta_i - \beta_s) Q_i},
\end{equation}
where, recall, in this case of a single collision the backward process coincides with the forward one, so that the distribution $\tilde{P}_\text{sc}$ in the denominator is simply $P_\text{sc}$. 
It is very important to emphasize, however,  that $P_\text{sc}(Q_i)$ is \emph{not} the marginal of $P(Q_1,\ldots,Q_N)$ (with the exception of $Q_1$). 
Notwithstanding, comparing with Eq.~(\ref{FT}), we  see that the full process in FIG.~\ref{fig:forward} is related to the single-collision processes according to 
\begin{equation}\label{sc}
    \frac{P(Q_1,\ldots,Q_N)}{\tilde{P}(-Q_N, \ldots, -Q_1)} = \frac{P_\text{sc}(Q_1)}{P_\text{sc}(-Q_1)}\ldots \frac{P_\text{sc}(Q_N)}{P_\text{sc}(-Q_N)}.
\end{equation}
This result is noteworthy, for the right-hand side is a product whereas the left-hand side is not. 
The full distribution $P(Q_1,\ldots,Q_N)$ \emph{cannot} be expressed as a product because the heat exchanges are, in general, not statistically independent. 
Notwithstanding, the ratio on the left-hand side of~(\ref{sc}) does factor into a product. 
The point, though,  is that this is not the product of the marginals, but of another distribution $P_\text{sc}$. 

One can also write a formula of the form~(\ref{sc}), but for only some of the heat exchanges. 
For instance, it is true that 
\begin{equation}
    \frac{P(Q_1,\ldots,Q_N)}{\tilde{P}(-Q_N, \ldots, -Q_1)} =
    \frac{P(Q_1,\ldots,Q_{N-1})}{\tilde{P}(-Q_{N-1}, \ldots, -Q_1)} \frac{P_\text{sc}(Q_N)}{P_\text{sc}(-Q_N)}.
\end{equation}
This kind of decomposition, however, depends crucially on the causal structure since it can only be done for future exchanges. For instance, we cannot write something involving $P(Q_2,\ldots, Q_N)$. 
The reason is that  $P(Q_1,\ldots, Q_{N-1})$ satisfies the fluctuation theorem~(\ref{FT}), but $P(Q_2,\ldots, Q_N)$ does not (since, after the first collision the system is no longer in a thermal state). 

\subsection{Information-theoretic formulation of the entropy production}

We define the entropy production associated with Eq.~(\ref{FT}) as 
\begin{equation}\label{Sigma}
\Sigma[\gamma_s] = \ln \frac{\mathcal{P}[\gamma_s]}{\tilde{\mathcal{P}}[\gamma_s]}= \sum\limits_{i=1}^N (\beta_i - \beta_s) Q_i[\gamma_s]. 
\end{equation}
The second equality is obtained using the detailed balance relation~(\ref{eq:DBR}). 
We emphasize that this is the entropy production associated with the choice of backward protocol used in Sec.~\ref{sec:back}, which may differ from other definitions in the literature~\cite{Esposito2010a,Santos2017a} (c.f. Ref.~\cite{Manzano2018} for a more detailed discussion). 

Alternatively, we can consider the entropy production from the perspective of the global trajectory $\gamma_{se}$ in Eq.~(\ref{gamma_se}). 
Using also that $Q_i[\gamma_s] = Q_i[\gamma_e]$, we can then write $\Sigma[\gamma_{se}]$ as 
\begin{IEEEeqnarray}{rCl}
    \Sigma[\gamma_{se}] 
    &=& \sum\limits_{i=1}^N \beta_i Q_i[\gamma_e] - \beta_s (E_{\alpha_N}^s - E_{\alpha_0}^s) \\[0.2cm]
    &=& \sum\limits_{i=1}^N \ln\frac{q_i(n_i)}{q_i(n_i')} + \ln \frac{p_0(\alpha_0)}{p_0(\alpha_N)}. 
\end{IEEEeqnarray}
The average entropy production may then be written as 
\begin{IEEEeqnarray}{rCl}
    \langle \Sigma[\gamma_{se}]\rangle &=& S(\sigma_N) - S(\sigma_0) + D(\sigma_N|| \sigma_0) \\[0.2cm]
    &&+ \sum\limits_{i=1}^N \bigg\{ S(\rho_i') - S(\rho_i) + D(\rho_i'|| \rho_i) \bigg\},
\end{IEEEeqnarray}
where $S(\rho) = - \tr(\rho \ln \rho)$ is the von Neumann entropy and $D(\rho'||\rho) = \tr(\rho'\ln \rho' - \rho' \ln \rho)$ is the quantum relative entropy. 
Here $\sigma_N$ is the final state of the system after the $N$ collisions. 

The important aspect of this result is that it depends only on \emph{local} changes in the ancillas. 
That is, all quantities refer to the local states $\rho_i'$ of each ancilla after the interaction. 
In reality, because the ancillas all interact with the system, they actually become indirectly correlated. 
These correlations are still represented indirectly in $\Sigma[\gamma_{se}]$, but they do not appear explicitly. 
This, ultimately, is a consequence of the choice of backward process that is used in the Jarzynski-W\'ojcik scenario~\cite{Jarzynski2004}.

\section{Conclusions}

To summarize, we have considered here the sequential heat exchange between a system and a series of ancillas. 
We assume all entities start in thermal state, but at possibly different temperatures. 
Moreover, all interactions are assumed to be described by thermal operations, which makes the identification of heat unambiguous. 
The main object of our study was the joint probability of heat exchange $P(Q_1,\ldots, Q_N)$ for a set of $N$ collisions. 
This object contemplates the correlations between heat exchange, a concept which to the best of our knowledge, has not been explored in the quantum thermodynamics community. 
We showed that $P(Q_1,\ldots, Q_N)$ satisfies a fluctuation theorem, which relates this joint distribution with single collision events. 
This result, we believe, could serve to highlight the interesting prospect of analyzing thermodynamic quantities in time-series and other sequential models. 

\section{Acknowledgments} 

The authors acknowledge E. Lutz and M. Paternostro for fruitful discussions. GTL acknowledges the financial support from the S\~ao Paulo funding agency FAPESP and from the Instituto Nacional de Ci\^encia e Tecnologia em Informação Qu\^antica. 
JPS would like to acknowledge the financial support from the CAPES (PNPD program).

\bibliography{ref.bib}

\end{document}